\documentclass[12pt]{aastex62}
\usepackage{booktabs, url, amsmath}
\def\lea{\mathrel{<\kern-1.0em\lower0.9ex\hbox{$\sim$}}}
\def\gea{\mathrel{>\kern-1.0em\lower0.9ex\hbox{$\sim$}}}

\graphicspath{{./}{./Figures/}{figures/}}
\received{}
\revised{}
\accepted{}
\submitjournal{ApJ}

\shorttitle{Cluster Formation Efficiencies }
\shortauthors{Chandar et al.}

\begin{document}

\doublespace

\title{A Tale of 3 Dwarfs: No Extreme Cluster Formation in Extreme Star-Forming Galaxies}
\author[0000-0003-0085-4623]{Rupali Chandar}
\affiliation{Ritter Astrophysical 
Research Center, University of Toledo, Toledo, OH 43606, USA}
\author{Miranda Caputo}
\affiliation{Ritter Astrophysical 
Research Center, University of Toledo, Toledo, OH 43606, USA}
\author[0000-0001-7413-7534]{Angus Mok}
\affiliation{OCAD University, Toronto, Ontario, M5T 1W1, Canada}
\author{Sean Linden}
\affiliation{Dept. of Astronomy, University of Massachusetts at Amherst, Amherst, MA 01003, USA}
\author{Bradley C. Whitmore}
\affiliation{Space Telescope Science Institute, 3700 San Martin Drive, Baltimore, MD, 21218, USA}
\author{Aimee Toscano}
\affiliation{Washington \& Jefferson College, Washington, PA 15301}
\author{Jaidyn Conyer}
\affiliation{University of Akron, Akron, OH 44304}
\author{David O. Cook}
\affiliation{Caltech/IPAC, 1200 E. California Boulevard, Pasadena, CA 91125, USA}
\author{Janice C. Lee}
\affiliation{Gemini Observatory/NSF’s NOIRLab, 950 N. Cherry Avenue, Tucson, AZ, 85719, USA}
\author{Leonardo Ubeda}
\affiliation{Space Telescope Science Institute, 3700 San Martin Drive, Baltimore, MD, 21218, USA}
\author{Richard White}
\affiliation{Space Telescope Science Institute, 3700 San Martin Drive, Baltimore, MD, 21218, USA}
\correspondingauthor{Rupali Chandar}
\email{rupali.chandar@utoledo.edu}
\begin{abstract}

Nearly all current simulations predict that outcomes of the star formation process, such as the fraction of stars that form in bound clusters ($\Gamma$),
depend on the intensity of star formation activity ($\Sigma_{\rm SFR}$) in the host galaxy.  
The exact shape and strength of the predicted correlations, however, vary from simulation to simulation.
Observational results also remain unclear at this time, 
because most works have mixed estimates made from very young clusters for galaxies with higher ${{\rm{\Sigma }}}_{\mathrm{SFR}}$ with those from older clusters for galaxies with lower ${{\rm{\Sigma }}}_{\mathrm{SFR}}$.
The three blue compact dwarf (BCD) galaxies ESO185-IG13, ESO338-IG04, and Haro11 have played a central role on the observational side because they have some of the highest known $\Sigma_{\rm SFR}$ and published values of $\Gamma$.
We present new estimates of $\Gamma$
for these BCDs in three age intervals (1-10~Myr, 10-100~Myr, 100-400~Myr), based on age-dating which includes H$\alpha$ photometry to better discriminate between clusters younger and older than $\approx10$~Myr.
We find significantly lower values for $\Gamma$ (1-10~Myr) than published previously.
The likely reason for the discrepancy is that previous estimates appear to be based on age-reddening results that underestimated ages and overestimated reddening
for many clusters, artificially boosting $\Gamma$ (1-10~Myr).
We also find that fewer stars remain in clusters over time, with $\approx15-39$\% in 1-10~Myr, 
$\approx5-7$\% in 10-100~Myr, and 
$\approx 1-2$\% in 100-400~Myr clusters.
We find no evidence that $\Gamma$ increases with $\Sigma_{\rm SFR}$.
These results imply that cluster formation efficiency does not vary with star formation intensity in the host galaxy. If confirmed, our results will help guide future assumptions in galaxy-scale simulations of cluster formation and evolution. 

\end{abstract}
\section{Introduction} \label{sec:intro}

Stars and clusters form in the densest regions of molecular clouds \citep[e.g.,][]{Lada03,McKee07}.
Star clusters are  therefore intricately tied to the star formation process.  
One key goal is to understand how conditions in the star-forming interstellar medium imprint onto young star and cluster populations.  These are typically characterized by two fundamental parameters, the star and cluster formation efficiencies.
The latter quantity, which is the fraction of stars that form in gravitationally bound clusters known as $\Gamma$, has played a prominent role in theoretical and observational studies over the past decade. 

The most striking theoretical claim about $\Gamma$ is that it increases systematically with the intensity of star formation, which can be traced using the star formation rate (SFR) per unit area or $\Sigma_{\rm SFR}$.
Early analytic frameworks relied on scaling relations to map 
cluster formation efficiency to the density spectrum of the ISM  and to $\Sigma_{\rm SFR}$ \citep{Kruijssen12}. 
In the intervening decade, new galaxy scale simulations 
have used a variety of approaches (including MHD, N-body, semi-analytic) and physical scales to predict the 
relationship between star and cluster formation.
The simulations account for different subsets of the relevant physics either self-consistently or with sub-resolution models to track the formation (and evolution) of entire cluster populations in dwarf \citep[e.g.,][]{Lahen19} 
and Milky-Way mass galaxies \citep[e.g.,][]{Grudic22}.
While they rely on a number of assumptions and prescriptions for (unresolved) star formation and stellar feedback, 
nearly all find that $\Gamma$ increases with $\Sigma_{\rm SFR}$ \citep[e.g.,][]{Kruijssen12,Li17,Li18,Pfeffer19,Lahen19,Grudic18,Grudic22}.

\begin{figure}[!ht]
	\centering
\includegraphics[width=\textwidth]{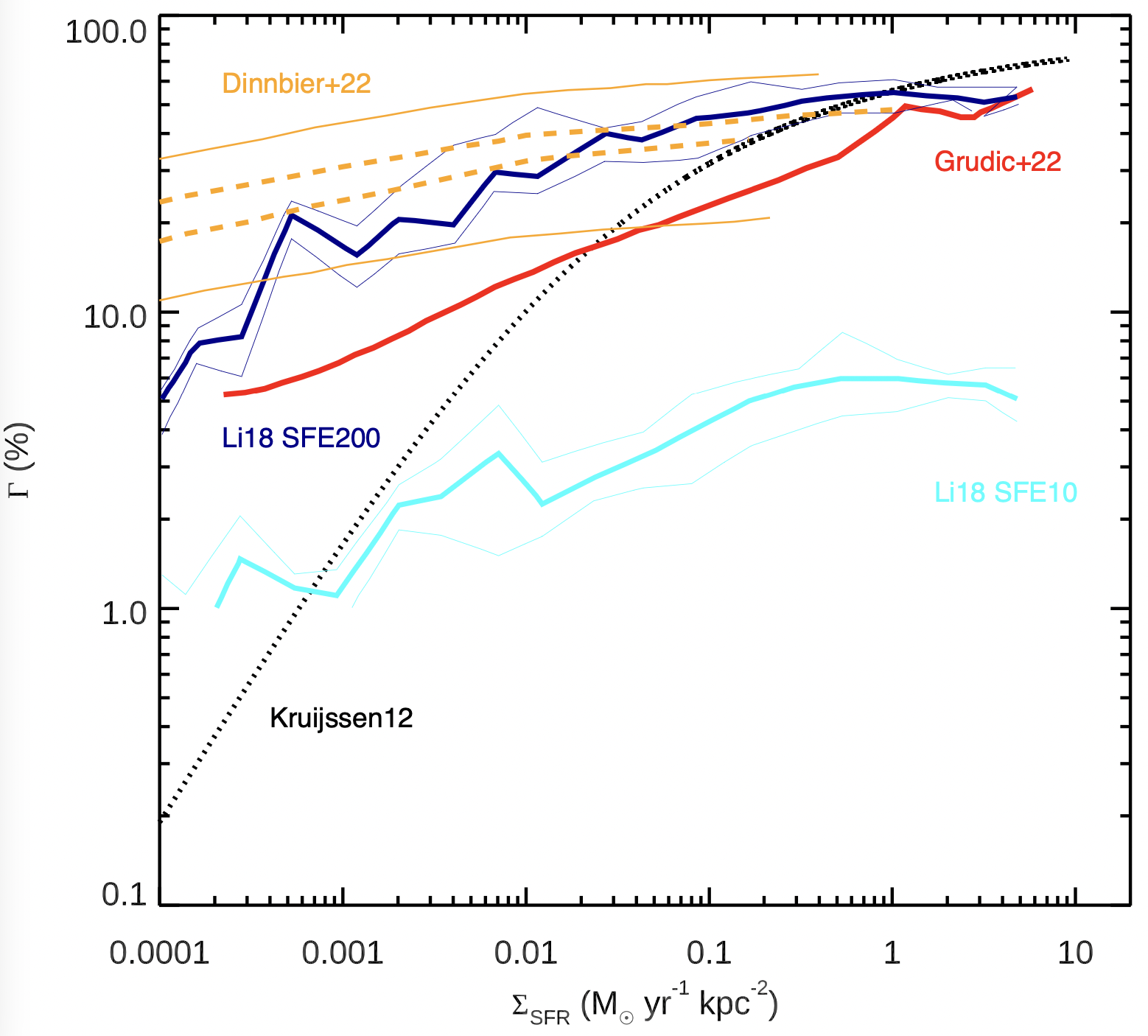}
	\caption{Predictions of the fraction of stars that form in bound clusters $\Gamma$ vs. $\Sigma_{\rm SFR}$, from five different simulations of cluster formation.  The predictions are reproduced from: \citet{Kruijssen12}, \citet{Li18}, \citet{Grudic22}, \citet{Dinnbier22}, as indicated. {\bf Main Point} --- the current generation of simulations predict an increase in cluster formation efficiency $\Gamma$ between a factor of $\approx100$   \citep{Kruijssen12} at the upper end, to 1.5 \citep{Dinnbier22} at the low end, over the range 0.001 to 1~$M_{\odot}~\mbox{yr}^{-1}~\mbox{kpc}^{-2}$ in $\Sigma_{\rm SFR}$.} \label{fig:GammaPredict}
\end{figure}

As shown in Figure~\ref{fig:GammaPredict}, the exact dependence of $\Gamma$ on $\Sigma_{\rm SFR}$ varies between simulations.
The early \citet{Kruijssen12} work predicted a $\sim100$-fold increase in $\Gamma$, from less than 1\% at low $\Sigma_{\rm SFR}$ ($\lea 0.001~M_{\odot}~{\rm yr}^{-1}~{\rm kpc}^{-2}$) to approaching 100\% at high $\Sigma_{\rm SFR}$ ($\gea 1~M_{\odot}~{\rm yr}^{-1}~{\rm kpc}^{-2}$).
A mapping of the GMCs in a FIRE-2 simulation onto a model cluster population predicts a more modest $\approx$factor of 10 increase in $\Gamma$ over the same $\Sigma_{\rm SFR}$ range \citep{Grudic22}, while other works predict shallower increases of factors $\sim2.5-4$, depending on the specific star formation and feedback parameters that are adopted \citep{Li18}. Recently, \citet{Dinnbier22} pointed out that changes in the assumptions (for e.g., the exact treatment of a star formation threshold and stellar feedback processes amongst others) can have a large impact on the predictions for $\Gamma$; they find a weak dependence (factor of $\sim1.5$) of $\Gamma$ on 
$\Sigma_{\rm SFR}$ in their preferred model, shown in orange in Figure~\ref{fig:GammaPredict}.

The observational picture is also unclear.  While many works have claimed that $\Gamma$ increases with $\Sigma_{\rm SFR}$ \citep[e.g.,][]{Goddard10,Cook12,Adamo15,Adamo20,Ginsburg18,Johnson16}, this trend arises, at least in part, because these works have compared estimates for $\Gamma$ calculated from different age intervals in different galaxies. 
At the low $\Sigma_{\rm SFR}$ end ($\approx0.001~M_{\odot}~{\rm yr}^{-1}~{\rm kpc}^{-2}$),   
estimates have mostly been made from $10-100$~Myr clusters.  Key examples are $\Gamma_{\rm 10-100~Myr}\sim5.8$\% and $\sim4.2$\% for the LMC and SMC, respectively \citep{Goddard10}, and $\sim 4-8$\% for M31 \citep{Johnson16}.  At the high $\Sigma_{\rm SFR}$ end ($\approx 1M_{\odot}~{\rm yr}^{-1}~{\rm kpc}^{-2}$),
estimates rely on $1-10$~Myr clusters. This includes results for the blue compact dwarf (BCD) galaxies that are the subject of this work, with published values of $\Gamma_{\rm 1-10~Myr}\sim50$\% for ESO338 and Haro11 \citep{Adamo11b}), and 26\% for ESO185 \citep{Adamo10}).  
It also includes the more massive, luminous infrared galaxies observed as part of the HiPeec survey (NGC~34, NGC~1614, NGC~3256, NGC~3690, NGC~4194, and NGC~6052), but which show no trend in $\Gamma_{\rm 1-10~Myr}$ over a factor of $\approx 10$ in $\Sigma_{\rm SFR}$ (see their  Figure~18; \citet{Adamo20}).
Because clusters naturally dissolve over time, mixing results from different age intervals can artificially imprint the claimed correlation \citep{Chandar17}.  

At least some of the confusion in the observational literature has likely been driven by the desire to include only bound clusters.  However, it is not possible to assess if clusters are bound (have negative total energy) or unbound, based on currently available information.
This means that extragalactic cluster catalogs include an unknown fraction of bound and unbound clusters at ages younger than $\lea 10$~Myr.  After this time, most remaining clusters are likely bound, although their observed age distributions suggest that they continue to lose mass and dissolve more-or-less continuously \citep[e.g.,][]{Fall05,Fall12,Chandar10a,Whitmore10}.  
With these competing considerations in mind, we will perform the same calculations to estimate the fraction of stars found in clusters in three age intervals
$1-10$~Myr,$10-100$~Myr, and $100-400$~Myr, 
and refer to the results as 
$\Gamma_{\rm 1-10~Myr}$, $\Gamma_{\rm 10-100~Myr}$, and $\Gamma_{\rm 100-400~Myr}$, respectively.

In this work we revisit estimates of the cluster formation efficiency for the BCD galaxies ESO185-IG13, ESO338-IG04, and Haro11, which have new WFC3 observations with the Hubble Space Telescope (HST) taken as part of the 'Clusters, Clumps, Gas, and Dust (CCDG) in Extreme Star-forming Galaxies' project (GO-15649, PI: Chandar).
Together with the luminous infrared HiPeec galaxies (which we will study in a future work), these BCDs hold up the high end of $\Sigma_{\rm SFR}$. 
 The rest of this paper is organized as follows. In Section~2, we summarize basic properties of our BCD sample, and present new cluster catalogs, including important updates on cluster age-dating.  At the end of this section we make a number of checks of cluster ages and identify some weaknesses in previous age-dating of these galaxies. The cluster mass functions are presented in Section~3, along with the results of maximum likelihood fits of the Schechter function to assess the shape at the high mass end.  In Section~4 we calculate the fraction of stars found in clusters, compare our results with those available for galaxies at the low $\Sigma_{\rm SFR}$ end, and discuss the implications.  Our main  results and conclusions are summarized in Section~5.
 We also include an Appendix, which presents the shape of the cluster mass functions (a step along the way to calculating $\Gamma$) in these BCDs for the first time, and perform a maximum likelihood fit to a Schechter function.

\begin{figure}[!ht]
	\centering
\includegraphics[width=\textwidth]{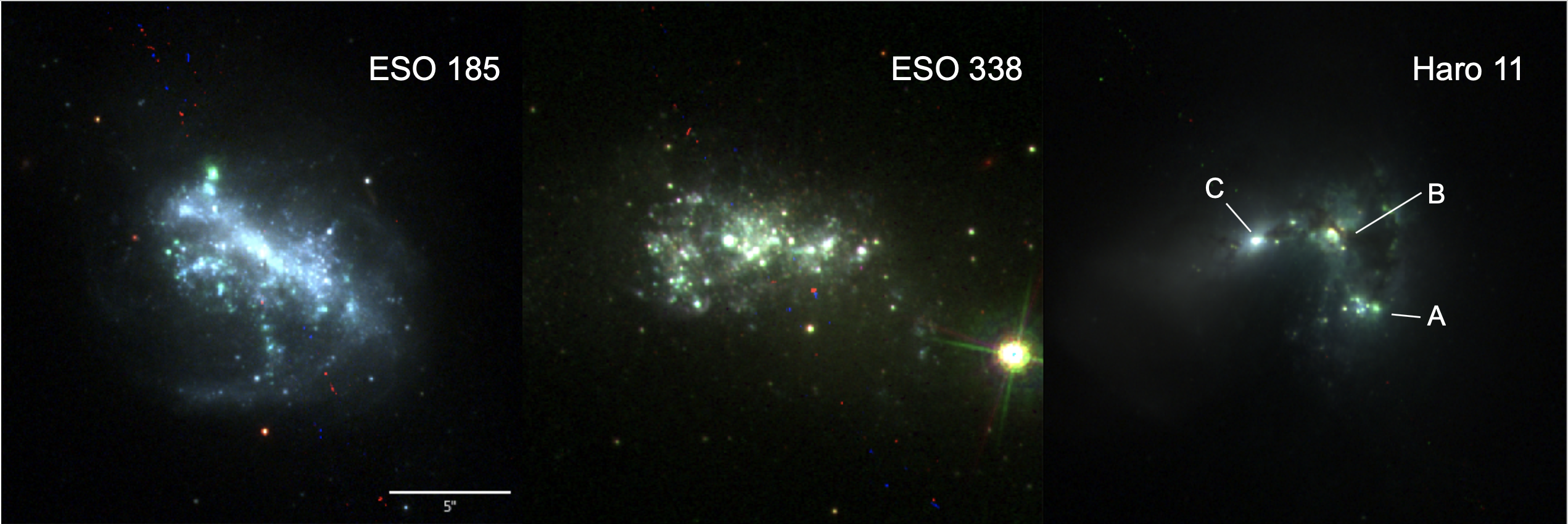}
	\caption{Three color B-V-I images of the Blue Compact Dwarf galaxies taken with the Hubble Space Telescope show hundreds of clusters with a broad range of colors (and hence ages).  Green point sources are H$\alpha$ emitters in ESO185-IG13 (left) and Haro11 (right).
	 The scale bar is $5\arcsec$ long.} \label{fig:sample}
\end{figure}

\section{Galaxy Sample and Cluster Catalogs}

\subsection{Galaxy Properties} \label{sec:gal}
\begin{table}[ht]
	\caption{Dwarf Galaxy Sample}\label{tab:sample}
	\centering
	\begin{tabular}{lccccccc}
	    \hline\hline
		Galaxy & Distance & Foreground  & SFR\tablenotemark{1}  & Area &  $\Sigma_{\rm SFR}$  \\
		Name & (Mpc) & E(B-V) (mag) & ($M_{\odot}~{\rm yr}^{-1}$) & (kpc$^2$) & $M_{\odot}~{\rm yr}^{-1}~{\rm kpc}^{-2}$)  \\ 		
		\hline \hline
 ESO~185-IG13  &  76 & 0.048 & 4.6 & 12.3 & 0.38 \\
ESO~338-IG04  & 38  & 0.076  & 2.3 & 2.0 & 1.15  \\ 
Haro11 & 82 & 0.010 & 17.4 & 10.2 & 1.7  \\ 
        \hline
	\end{tabular}
\tablenotetext{1}{The star formation rates have been converted from an assumed Salpeter to Chabrier IMF by dividing by 1.38.}
	\end{table}
	\par 
Color images of ESO185, ESO338, and Haro11 taken in the BVI filters with the HST are shown in Figure~\ref{fig:sample}.  ESO185 has a bright, bar-like structure crossing the center, and tidal features indicating the system experienced a recent merging event. Recent star and cluster formation is more evenly spread throughout ESO338 than in ESO185.  BCDs in general have low internal extinction \citep[e.g.,][]{Mas-Hesse99}.
Haro11, which has three bright 'knots' (labeled A, B, C), is the only one of the three galaxies to show limited dusty regions. We adopt the same distances as \citet{Adamo10,Adamo11a} to each galaxy, and compile this information in Table~1.

We adopt H$\alpha$-based estimates of the star formation rate, since this tracer is sensitive to only the most recent ($\approx 10$~Myr) star formation \citep[e.g.,][]{Kennicutt12}, and a close match to the $1-10$~Myr interval used to determine $\Gamma$.  
\citet{Adamo10,Adamo11a} and \citet{Hayes07} (Haro11) used the \citet{Kennicutt98a} calibration to convert total H$\alpha$ luminosity to SFR for each galaxy.  However, this calibration assumes a Salpeter stellar IMF, whereas we adopt a \citet{Chabrier03} IMF when we estimate cluster masses in Section~2.4.\footnote{\citet{Adamo10,Adamo11a}  adopted a Kroupa IMF (similar to the Chabrier IMF assumed here) when estimating cluster masses, but used H$\alpha$-based star formation rates that assumed a Salpeter IMF.}  We therefore divide the published H$\alpha$-based SFRs by a factor of 1.38 to account for the difference between the Salpeter and Chabrier IMFs.  The corrected SFRs are: 
$4.6~M_{\odot}~\mbox{yr}^{-1}$ (ESO185),
$2.3~M_{\odot}~\mbox{yr}^{-1}$ (ESO338), and
$17.4~M_{\odot}~\mbox{yr}^{-1}$ (Haro11), and are
compiled in column~4 of Table~1.  

In BCDs, star formation is confined to an area that 
is $\approx100$ times smaller than a typical spiral disk.  We adopt the same area (in kpc$^2$) as \citet{Adamo10,Adamo11a}  (listed in column~5 of Table~1).  The moderate-to-high rates of star formation contained within a small area lead to very high values of $\Sigma_{\rm SFR}$: 0.38 (ESO185), 1.15 (ESO338), and $1.7~M_{\odot}~{\rm yr}^{-1}~{\rm kpc}^{-2}$ (Haro11), compared with $\sim0.01~M_{\odot}~{\rm yr}^{-1}~{\rm kpc}^{-2}$ for spiral galaxies like M51 and M83 \citep[e.g.,][]{Adamo15,Chandar17}.  Dwarf irregular galaxies like the LMC and SMC have $\Sigma_{\rm SFR}$ which are lower than spirals by a factor of $\approx10$ \citep[e.g.,][]{Goddard10, Cook19, Whitmore20}, and hence lower than the BCDs studied here by a factor $\approx1000$.

There are several inherent sources of uncertainty in the SFR estimates, including measurements of flux, variations in the SFR, the lower metallicity of our galaxies, and the leakage of Lyman continuum photons from the parent galaxy.\footnote{Haro11 is in fact, well-known to leak these ultraviolet photons.}
Following \citet{cook23} we adopt a 25\% uncertainty in the SFR estimates.



\subsection{Data, Detection, and Photometry}

\begin{table}[ht]
	\caption{WFC3 Filters and Exposure Times}\label{tab:filters}
	\centering
	\begin{tabular}{lcccccccc}
	    \hline\hline
		Galaxy & NUV (F275W)  & U (F336W) & B (F438W) & V (F555W) & I (F814W) & H$\alpha$ (F665N) & P$\beta$ (F130N) & H (F160W)  \\ \hline \hline
ESO~185-IG13 & 1935 & ... & 1890$^{1}$ & 2680 & 1800 & 1905 & 800 & 1600  \\
ESO~338-IG04 & 1965 & ... & 1500 & 830$^{2}$ & 1350 & ... & 1802 & 2400  \\
Haro11 & 1920 & 1332 & 680$^{1}$ &  1740 &  1740 & 1800 & 800 & 1600 \\
        \hline
	\end{tabular}
	\end{table}

 HST has imaged ESO185, ESO338, and Haro11 in broad- and narrow-band filters covering the near-ultraviolet to the near-infrared.  The observations used in this work are a mix of new and archival images, with the new data taken as part of our CCDG program (GO-15649; PI: Chandar) with the WFC3 camera.  We primarily use observations in the NUV (F275W), U (F336W) B (ACS/F435W or WFC3/F438W), V (WFC3/F555W or ACS/F550M), H$\alpha$ (F658N), and I (F814W) filters, but checked for highly reddened sources in Haro11 using the P$\beta$ (F130N) and H (F160W) images as well. The lower resolution of the near-infrared images makes photometry in these bands less useful for age-dating.
 ESO185 and ESO338 have not been imaged in the U band, and ESO338 also does not have H$\alpha$ imaging.
\par 
  Individual exposures are processed through the standard Pyraf/\linebreak[0]{}STSDAS CALACS or CALWFC3 software, 
then aligned and drizzled onto a common grid to create one image for each filter using the DRIZZLEPAC software package.  The V band image is used as the reference in each case, with a pixel scale of $0\farcs04$ (the native WFC3 pixel scale), and astrometry calibrated with {\bf Gaia} DR2 sources \citep{gaiadr2}. 
Sky subtraction is performed during the drizzling process and the output images are all oriented with North up and East to the left.  The final FITS files are in units of
electrons per second.
\par 
At the distances of the three BCDs in our sample, stellar clusters appear essentially as point sources.  We run the DAOFIND detection algorithm on 
the $V$-band image in each galaxy with a $3\sigma$ detection limit.
Aperture photometry was performed in a two-pixel radius for all sources in all broad-band filters, with an annulus with radii between 7 and 9~pixels to determine the background level. These apertures were selected because they minimize the scatter in the measured colors of clusters compared with predictions from stellar evolutionary models.  We convert the apparent magnitudes to the VEGAmag system by applying the following zero points: 22.640 (WFC3/F275W), 23.526 (WFC3/F336W), 25.764 (ACS/F438W) , 25.832 (WFC3/F555W), and 25.518 (WFC3/F814W).
Filter-dependent aperture corrections  
were determined from isolated sources and applied to obtain the total apparent magnitude for each cluster.
Photometric errors include uncertainties in both source counts and  background levels.

\begin{figure}[!ht]
	\centering
\includegraphics[width=\textwidth]{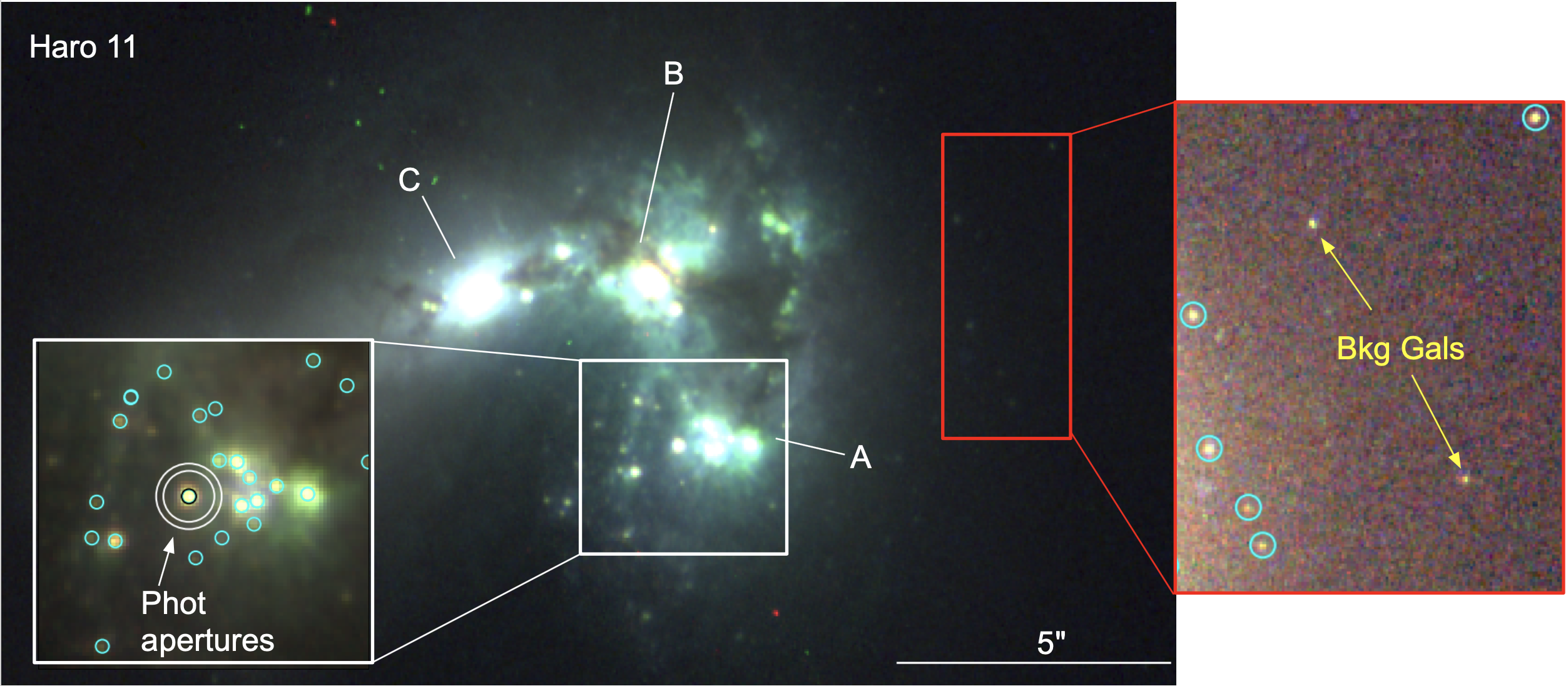}
	\caption{{Three color B-V-I image and zoom in to a crowded, actively star-forming region dominated by young, $\sim15-20$~Myr clusters (outlined in white), and another region towards the outskirts dominated by clusters older than 100~Myr (outlined in red). In the star-forming region we only show the locations of clusters brighter than 23.5~mag in the V band, since it is hard to show bright and faint clusters in the same stretch.  
 The 2 pixel aperture used to perform photometry, and the 7 and 9~pixel annuli used for background subtraction are also shown.  Two background galaxies that were excluded from our cluster catalog are identified in the outer region.}} \label{fig:detections}
\end{figure}

The narrow-band H$\alpha$ images, which trace warm ionized gas, sometimes have a very different morphology than the stellar emission that dominates the broad-band images. Feedback from recently formed massive cluster stars quickly push the surrounding ionized gas into bubbles and partial shells with a range of radii.  We therefore do not include background subtraction in the aperture photometry performed on the non-continuum subtracted, narrow-band images.

We select clusters to have a measured V-band magnitude brighter than 27.0~mag, and to have a concentration index between 1.0 and 2.4 (concentration index is the difference in aperture magnitudes in the V-band determined using a 2 pixel
and a 0.5 pixel radius).  The upper limit selected for the measured concentration index eliminated detections of diffuse regions within the galaxies, while the lower limit eliminated a handful of cosmic ray residuals.
The main sources of contamination that remain after these selection criteria are foreground stars and background galaxies. 
We performed a final visual inspection of the detected sources in all three galaxies to remove remaining contaminants.  ESO185 and Haro11 each contain objects which appear to be nuclei or nuclear clusters (a single object at the dynamical center of ESO185, and three very bright point-sources in Haro11).  We do not include these sources as part of the cluster analysis.
The final cluster catalogs contain $213/596$ (ESO185), $300/399$ (ESO338), and $180/536$ (Haro11) candidate clusters brighter than $m_V=24.5/27.0$; 
clusters with $m_V$ of 24.5~mag and brighter have a median uncertainty of $\approx0.2$~mag in the NUV filter.

\subsection{Color-color Diagrams and Constraints on Reddening}

\begin{figure}[!ht]
	\centering
\includegraphics[width=\textwidth]{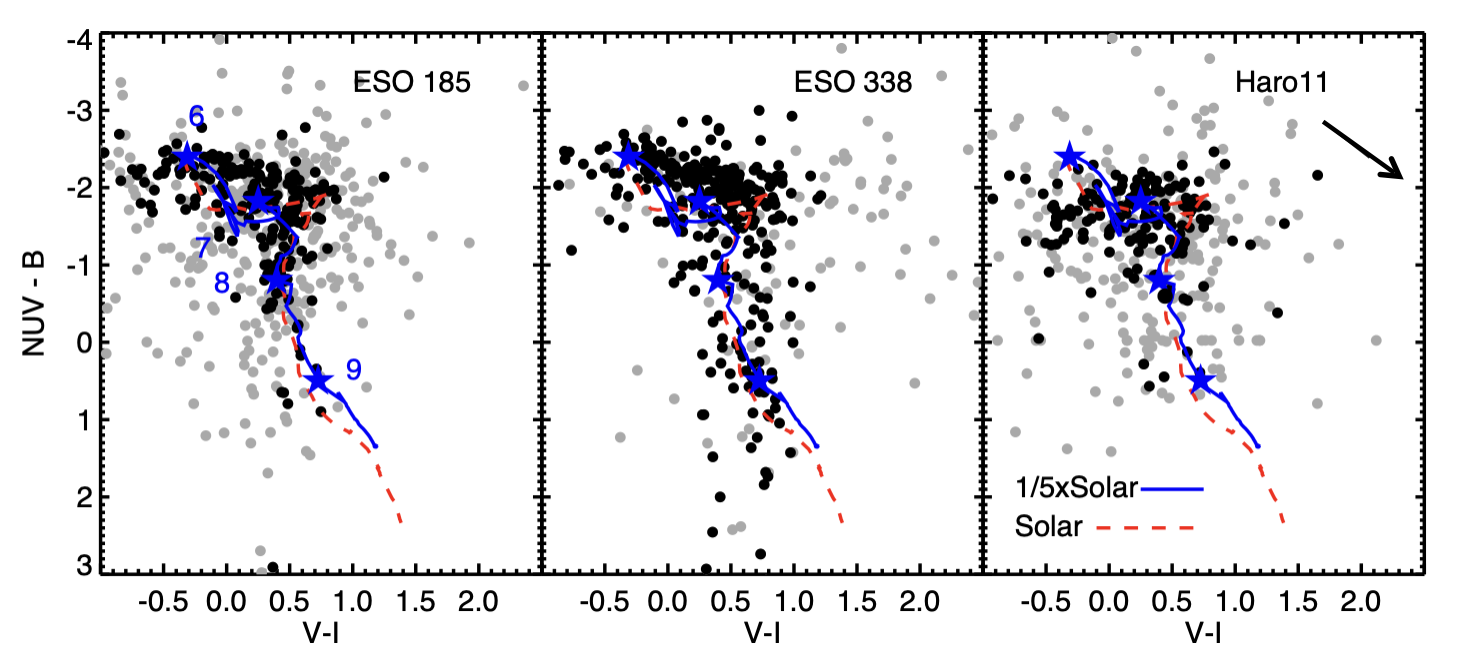}
	\caption{NUV-B vs. V-I  color-color diagrams for clusters detected in our blue compact dwarf galaxies.  The solid lines in each panel show the $1/5\times$solar metallicity predictions from the \citet{Bruzual03}  models (hereafter BC03), appropriate for these dwarf galaxies.  The dashed lines show predictions from the solar metallicity BC03 model.  The arrow in the right panel shows the direction that reddening will move cluster colors. All clusters with photometry in the NUV, B, V, and I bands are shown as the gray points, and clusters brighter than $m_V \approx 24.5$~mag are shown as the black points.  {\bf Main Point}---The measured cluster colors follow the model predictions quite well, indicating that overall, reddening does not significantly affect cluster colors. } \label{fig:colcol}
\end{figure}

The NUV-B vs. V-I color-color diagrams of the cluster populations in the three dwarf galaxies are shown in Figure~\ref{fig:colcol}.  All detected clusters in ESO 185 (left panel), ESO 338 (middle panel), and Haro~11 (right panel) are shown as gray circles, and clusters brighter than $m_V$ of 24.5~mag are shown as the black circles.  The cluster colors have been corrected for the foreground reddening values given in Table~1.  The arrow in the right panel shows the direction that additional reddening would move a cluster, assuming a Milky Way-type extinction law \citep{Fitzpatrick99}.
Each panel also shows predictions from the \citet{Bruzual03} stellar population models\footnote{We previously tested several different models, and found that overall, the \citet{Bruzual03}, hereafter BC03, population synthesis models best match the observed colors of young star clusters \citep[e.g.,][]{Chandar10b,Turner21}.} 
for 1/5$\times$solar  (blue solid line) and solar metallicity (red dashed line).  The models predict the color (and luminosity) evolution of clusters starting soon after their birth at 1~Myr (upper-left), through the ages of globular clusters around $\sim12$~Gyr in the lower-right, and do not include any contribution from nebular line emission.  

We see that in aggregate, the cluster colors, especially the black points brighter than $m_V$ of 24.5~mag, closely follow along the predicted model tracks, 
except for a handful of sources in Haro11 which are to the right of the BC03 models.  This suggests that the overall cluster populations in these BCDs do not experience much reddening due to dust.  
	After examining colors in different locations, we find that only clusters in the region around and to the west of Knot~B in Haro 11 have any appreciable reddening. We perform age-dating experiments which show that a maximum E(B-V)$\approx0.5$~mag ($A_V\approx1.5$~mag) (when H$\alpha$ photometry is included in the SED fitting) is sufficient to successfully age-date the handful of reddened young clusters in this area. 
 
\begin{figure}[!ht]
	\centering
\includegraphics[width=\textwidth]{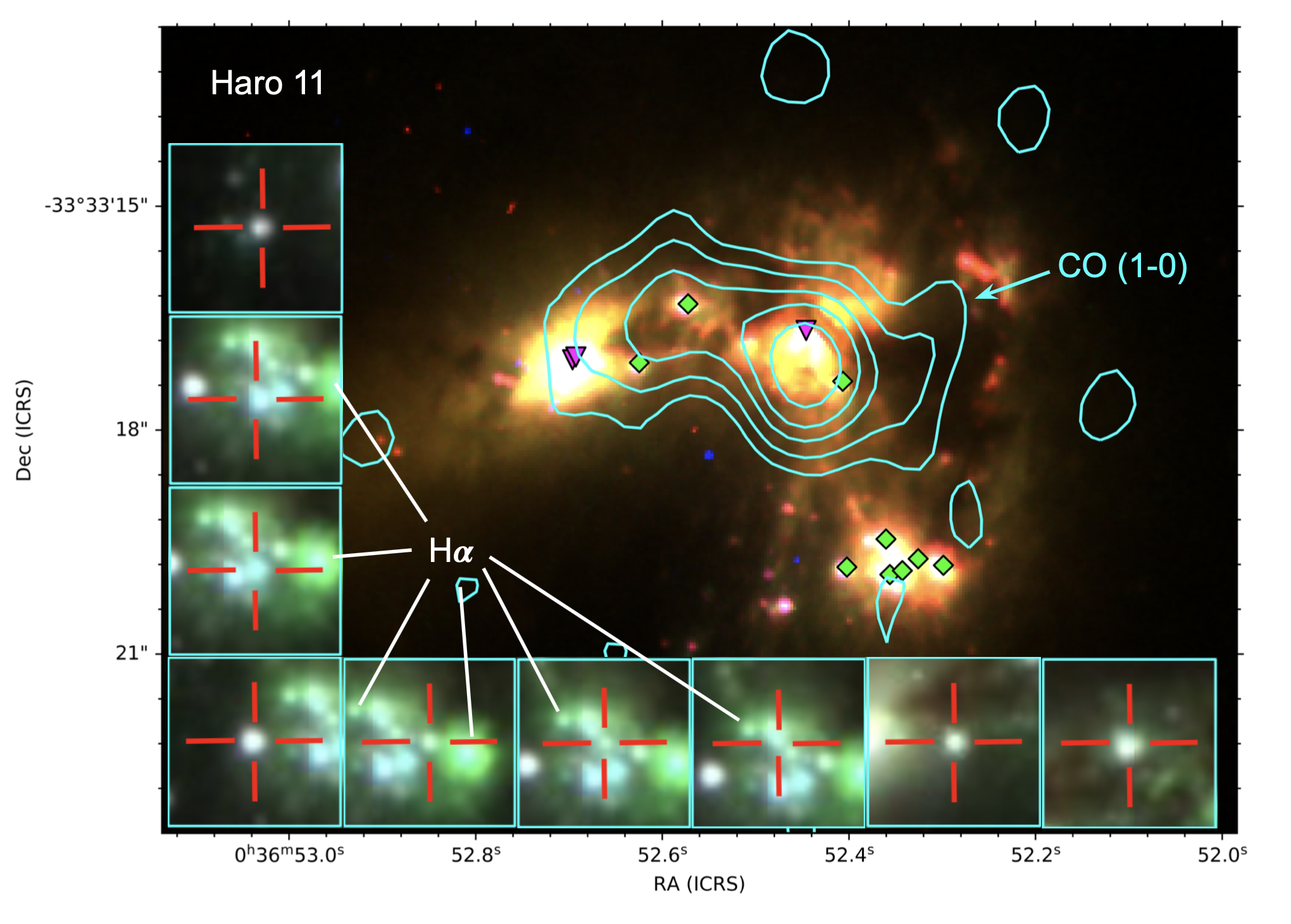}
	\caption{Contours of CO (1-0) are plotted on top of a HST NUV-B-V three-color image of Haro11.
The strongest CO emission is approximately coincident with Knot B, one of three nuclear regions (identified by purple triangles) that have been removed from the cluster sample. The second strongest peak is coincident with a bright, optical cluster.
 There is also weak CO emission in the dusty region to the west of these strong CO emitting regions, and in the region to the south where the most massive young clusters have formed. 
 The 9 most massive clusters younger than $\sim15-20$~Myr are identified by green diamonds, along with {\bf 1.4$\arcsec$ $\times$ 1.3$\arcsec$} postage stamps showing what they look like in the B-V-I filters.  None of these 9 clusters have H$\alpha$ emission (seen as green emission in the postage stamps) directly associated with them, although there are a couple of strong H$\alpha$ emitting sources near some of these clusters. 
 } \label{fig:haro11CO}
	\end{figure}

We also produced CO(1-0) moment-0 emission line maps of Haro11 from archival data taken by the Atacama Large Millimeter Array (ALMA; 2013.1.00350.S).
These contours are shown in Figure~\ref{fig:haro11CO} superposed on a color $HST$ image.  The strongest peaks of CO emission are coincident with bright optical region, not with the dustiest parts of the galaxy where a handful of reddened young clusters are found (although we do find lower levels of CO emission from these dusty regions).  We also see little CO in the region towards the south (Knot A) which has recently formed many of the most massive clusters.  No P$\beta$-emitting sources are revealed in this dusty area, indicating that either clusters are not currently forming in these regions, or if they are they remain deeply buried, with $A_V \gea 7-10$~mag.

\section{Cluster Age Dating}

\subsection{Method and Checks}

\begin{figure}[!ht]
	\centering
    \includegraphics[width=\textwidth]{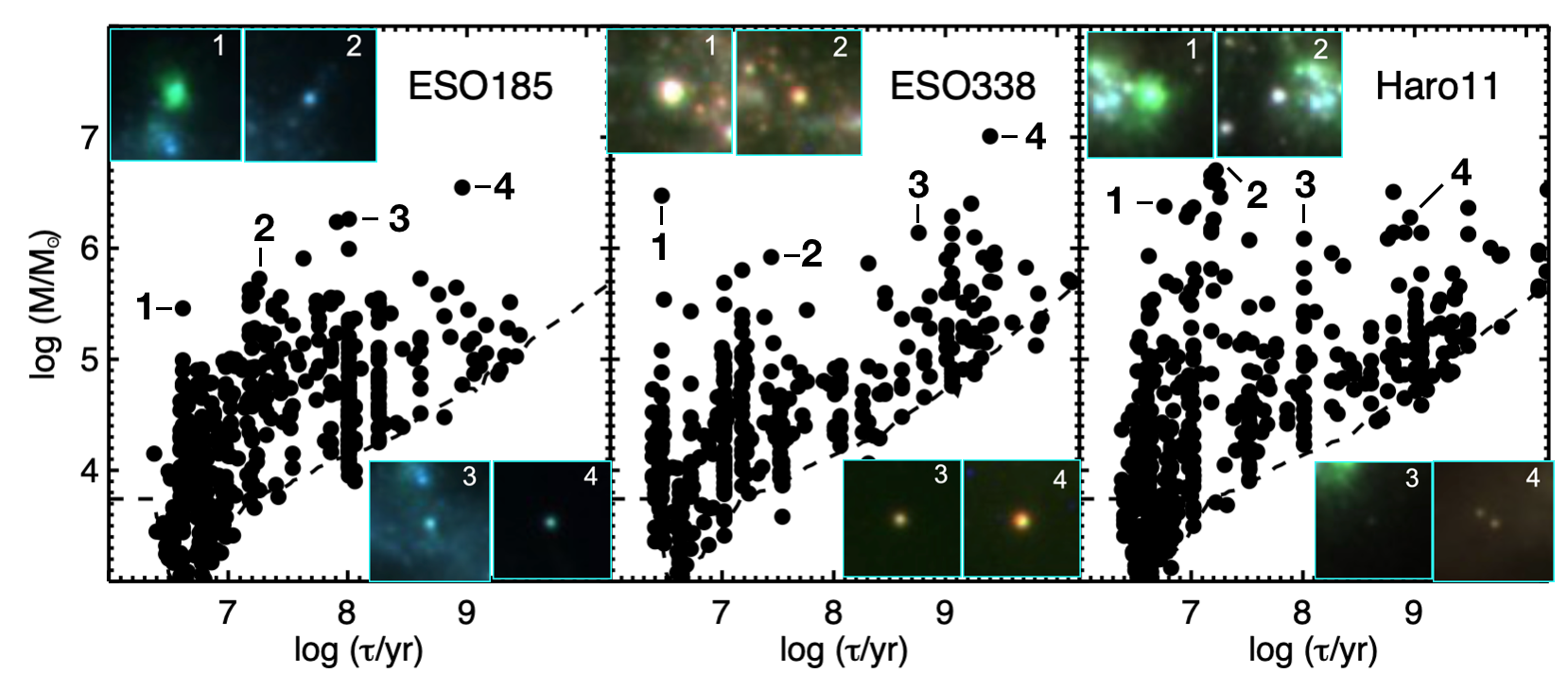}
	\caption{Cluster age-mass diagrams for our sample galaxies.  The dashed line represents $M_V=-8$~mag, the approximate brightness limit for individual stars, and shows that our sample does not include older clusters at lower masses.  The cluster population in each dwarf galaxy has a unique mass-age distribution.  Postage stamp B-V-I images of different age cluster are shown, as indicated.  The very young $3-6$~Myr clusters in ESO185 and Haro11 (for example those labelled '1') have strong H$\alpha$ emission associated with them, while the $\approx15-20$~Myr clusters (e.g., labelled '2') do not.} \label{fig:Mt}
\end{figure}

To estimate the mass, age, and reddening of each cluster in the three BCD galaxies, we follow the general SED-fitting technique described in \citet{Chandar10b},
but with an updated treatment of the reddening.
We estimate the age ($\tau$) and extinction ($A_V$) for each cluster by performing a least $\chi^2$ fit comparing the observed cluster magnitudes with predictions from the $1/5\times$solar BC03 population model.  The grid runs over ages from log~($\tau/\mbox{yr})=6.0$ to 10.2, and from E(B-V)$=0.0$~mag up to a maximum value.  Most works adopt a single maximum E(B-V) value for the entire cluster population during SED fitting, based on the idea that sight lines to some individual clusters may be highly reddened.
For example a maximum E(B-V) of $1.5$~mag was adopted by the LEGUS \citep{Calzetti15} and PHANGS  \citep[e.g.,][]{Turner21,Lee22} cluster surveys, and by \citet{Adamo10,Adamo11a} for the clusters in ESO185-IG13, ESO338-IG04, and Haro11.  However, adopting a single, large value like 1.5~mag for the maximum E(B-V)  
 can result in serious problems in the estimated ages for clusters older than $\approx10$~Myr, as explored and discussed in detail by \citet{Whitmore20,Whitmore23}. 

\par 
Here, we use a more direct approach and allow the adopted maximum E(B-V) value used during SED fitting to vary based on observations and inferences about the clusters themselves. 
As noted earlier,  color-color diagram of clusters in ESO185 and ESO338 (Figure~\ref{fig:colcol}) 
show very little reddening towards nearly the entire cluster population.
We therefore adopt a maximum E(B-V)$=0.1$~mag for all clusters in these two galaxies during SED fitting,  but also check the impact of adopting a maximum value of 0.5~mag and 1.0~mag in the age-dating.  While most clusters in Haro11 also follow the BC03 models, there are a handful in the dusty region near Knot~B that fall off the BC03 model tracks along the reddening vector.  These clusters have H$\alpha$ emission, and therefore are $\approx3-6$~Myr old. If we compare their observed colors with those predicted for these young ages, these clusters should have E(B-V)$\approx0.5$~mag.  For Haro11, we therefore adopt a maximum reddening of E(B-V)$=0.5$~mag for the reddened clusters near knot B, and E(B-V)$=0.1$~mag for the rest.
\par 
The best-fit values of $\tau$ and $A_V$ minimize the statistic: $\chi^2(\tau, A_V ) = \sum_{\lambda}~W_{\lambda} (m^{\mbox{obs}}_{\lambda}- m^{\mbox{mod}}_{\lambda})^2$ where $m^{\mbox{obs}}$ and $m^{\mbox{mod}}$ are the observed and model magnitudes respectively, and the sum runs over all available broad- and narrow-band filters NUV, U, B, V, H$\alpha$, and I. The weight factors in the formula $\chi^2$ are taken to be $W_{\lambda}=[\sigma_{\lambda}^2 +(0.05)^2)]^{-1}$, where $\sigma$ is the photometric uncertainty.  The mass of each cluster is estimated from the observed V band luminosity, corrected for extinction, and the (present- day) age-dependent mass-to-light ratios ($M/L_V$) predicted by the models, assuming the distances compiled in Table~1.  We note that correcting for reddening internal to a cluster can significantly boost it's estimated mass in some cases. For example, a reddening of E(B-V)$\approx0.5$~mag corresponds to $A_V\approx1.6$~mag, which would boost the intrinsic luminosity and hence cluster mass by a factor of more than 4 for a given age and mass-to-light ratio.
\par 
The largest source of uncertainty in the age estimates comes from uncertainties in the photometry,  particularly in the NUV and $U$ band measurements for the fainter clusters. Our method typically produces uncertainties at the level of log~$(\tau/\mbox{yr}) \approx0.3$ ($\approx$factor of 2); see e.g., \citet{Chandar10b}.
The uncertainties in the mass estimates are tied to those in the ages and to a lesser degree to the best fit reddening.
The clusters in our study are sufficiently massive that stochastic fluctuations should not have much impact on their colors.
\par 
Our method, which allows for a flexible maximum $A_V$ in the SED fitting based on the amount of reddening in a given region, improves age results for clusters in Haro11.  Specifically, the bright cluster located at RA$=$00:36:52.41 and DEC$=$-33:33:16.58 in Knot B (see Figure~\ref{fig:haro11CO}, has red broad-band colors, but some H$\alpha$ emission right on top of and next to it.  Our age-dating method (which includes H$\alpha$ for Haro11) finds log~age$\approx9$ when a maximum E(B-V) of 0.1~mag is imposed, but log~age$\approx6.98$ when a maximum E(B-V) of 0.5~mag is assumed instead.  
The final cluster age-mass results are shown in Figure~\ref{fig:Mt}.
\par 
We perform a number of checks of our age-dating results.  One important take-away is that our method returns estimated ages younger than 10~Myr for clusters with obvious H$\alpha$ emission regardless of whether we adopt a maximum E(B-V) of 0.1, 0.5, or 1.0~mag in the SED fitting; the handful of highly reddened, very young clusters in Haro~11 discussed in Section~3.1 are the only exceptions, and require an E(B-V) of at least 0.5~mag.  Adopting a maximum E(B-V) of 0.5 or 1.0~mag affects only $\approx7$\% of the sample, predominantly clusters fainter than $m_V\sim 25$~mag, and systematically changes their estimated ages from $\sim30-500$~Myr with little reddening to just under 10~Myr with moderate reddening.  It is hard to draw firm conclusions about the actual ages of these faint clusters, although most of them are in regions with little dust, suggesting they are likely faded, older clusters.

We visually inspected every cluster with $M\geq 10^5~M_{\odot}$ in color images.
In Figure~\ref{fig:Mt} we include postage stamp images of 4 massive clusters in each galaxy as illustrative examples.  In each case, clusters designated '1' highlight one cluster younger than 10~Myr, which are typically the brightest H$\alpha$ emitting source in it's parent galaxy.  Clusters designated as '2' all have best fit ages between $\sim 15-30$~Myr, with blue colors but no associated H$\alpha$ emission.  Clusters designated as '3' and '4' are somewhat older, with redder colors (but no associated dust) and no H$\alpha$ emission.
\par 
As mentioned earlier, Haro11 is the only BCD in our sample which has appreciable dust.
Our results indicate that some massive clusters in Haro11 have ages $\approx3-6$~Myr (e.g., cluster '1' in the right panel of Figure~\ref{fig:Mt}), while other massive clusters have ages $\approx15-20$~Myr (e.g., cluster '2'). One potential concern is that the clusters assigned ages of $\approx15-20$~Myr are actually younger, but moderately reddened. An inspection of all clusters more massive than $10^5~M_{\odot}$ confirms that those with estimated ages of $\approx15-20$~Myr have no associated H$\alpha$ emission or visible dust, 
as seen in the postage stamps included in Figure~\ref{fig:haro11CO}.   Therefore the age-dating method appears to be able to correctly differentiate between and age-date clusters formed $\approx3-6$~Myr ago from those formed $\approx15-20$~Myr ago.

\subsection{Comparison with Previous Works}

The cluster populations of ESO185, ESO338, and Haro11 were previously studied by Adamo et al. (2010, 2011) based on earlier $HST$ imaging, some with the WFPC2 camera.   While we do not have access to their catalogs, we are able to make some global comparisons.  Our cluster catalogs have at least twice as many clusters as the earlier ones, likely driven by the shallower observations in the three bluest filters that were available to Adamo and collaborators more than a decade ago. 
They presented mass-age and reddening-age diagrams for clusters in ESO185 and Haro11, allowing us to make some 
comparisons of age-dating results for these two galaxies.
The ages and masses of clusters in Haro11 were more recently determined by \citet{Sirressi22} based on photometry made from the same observations used here, although no estimate of $\Gamma$ was made.

There are some important differences in the results between our work and that of \citet{Adamo11a,Adamo11b} for ESO185 and Haro11.
For ESO185, \citet{Adamo11a} find $\sim25$ clusters with E(B-V) between $0.5-1.2$~mag (see their Figure~10), despite the low dust content in this dwarf galaxy. They find 10 clusters  with ages log$\tau < 7.0$ and masses $M >10^5~M_{\odot}$, whereas we find a single cluster in this age-mass range ('1' in the left panel of Figure~\ref{fig:Mt}). In fact, \citet{Adamo11a} find that a majority of clusters are younger than log$\tau < 7.0$, and only 9 clusters in their sample have estimated ages between log~$\tau = 8.0-9.0$ (see their Figure~11).  Our age-dating meanwhile, finds 130 (more than 10 times) as many clusters with ages between log~$\tau = 8.0-9.0$.

We find more subtle differences in the age-dating results for Haro11.  
\citet{Adamo11a}  find $\sim15$ clusters with E(B-V)$\geq 0.5$~mag, and a dominant, very young $\sim3-5$~Myr cluster population with masses approaching $10^7~M_{\odot}$, and almost no clusters older than log$\tau=7.6$.  We also find very massive clusters younger than log$\tau < 7.0$, although the most massive cluster in our sample (designated as '1' in the right panel of Figure~\ref{fig:Mt}) is $\approx1/2$ the mass of theirs, and we only find a handful of clusters are best fit with E(B-V)$\approx0.5$~mag.  The most notable difference is the estimated ages of the most massive clusters in the two works---here we find ages of $\sim15-20$~Myr from our broad-band plus H$\alpha$ photometry, while \citet{Adamo11a} find $\approx3-4$ Myr (but did not have H$\alpha$ photometry).   
\citet{Sirressi22} meanwhile, find 6 clusters younger than 10~Myr with estimated masses $\approx10^7~M_{\odot}$, significantly more than in \citet{Adamo11b} or this work.  We suspect this is due to their assumption of a high maximum E(B-V) of 1.5~mag during age-dating, and details of their SED fitting processs.
As described earlier, we have visually inspected all of our $\sim15-20$~Myr clusters and believe our age dating is correct, since none of these clusters have H$\alpha$ emission associated with them.

We believe there are two main reasons for the differences described above: (1) we include  H$\alpha$ photometry directly in the SED fits, and (2) \citet{Adamo10,Adamo11a}  adopted a significantly higher maximum E(B-V) of 1.5~mag in their age-dating procedure, compared with the 0.1~mag we adopt for nearly all clusters.
SED fits which only include broad-band colors 
struggle to break the age-reddening degeneracy and can result in a number of systematic, 'catastrophic' failures in age-dating, with many older clusters with little reddening erroneously fit by a combination of young age and high reddening (e.g., Whitmore et al. 2020; Whitmore et al. submitted; Floyd et al., in prep). Including H$\alpha$ photometry directly in the SED fits helps break the age-reddening degeneracy in many cases \citep[e.g.,][]{Chandar16,Ashworth17}.    
We suspect that the high maximum reddening adopted by \citet{Adamo10,Adamo11b}  for these low metallicity, low dust content dwarf galaxies resulted in a number of clusters erroneously fit by a combination of too young age plus too-high reddening.  This essentially moved clusters in the age-mass diagram from ages near clusters designated with a '2' and/or '3' in Figure~\ref{fig:Mt} to ages near those designed with a '1'.
We discuss the impact this common age-dating problem 
may have had on previous estimates of $\Gamma$ in Section~4.4.

\section{New Measurements and Implications of the Cluster Formation Efficiency}

Despite nearly a decade and a half of observational work, the relationship between 
$\Gamma$ and ${{\rm{\Sigma }}}_{\mathrm{SFR}}$ (or SFR)  remains unclear in nearby galaxies.
Estimates of $\Gamma$ in galaxies with both high and low $\Sigma_{\rm SFR}$ in particular, are needed 
to solidify the observational picture and to compare with current simulations.

\subsection{Estimates of $\Gamma$}

The three BCDs studied in this work have among the highest published estimates of $\Gamma_{\rm 1-10~Myr}$, and also among the highest $\Sigma_{\rm SFR}$ for galaxies within 100~Mpc.  In Section~2 we noted a key shortcoming in the previous estimates of $\Gamma_{\rm 1-10~Myr}$ for these galaxies, that the estimated ages and masses for a subset of the clusters are likely incorrect, as a result of specific (common) assumptions made during SED fitting.  In this section we calculate new values of $\Gamma$ (in three age intervals), and compare with results published for galaxies with low $\Sigma_{\rm SFR}$.

We follow the exact same methodology as Cook et al. (2023) and estimate $\Gamma_{\rm 1-10~Myr}$ to be the mass of stars born in ($1-10$~Myr) clusters divided by the total stellar mass formed over this time period.   
The total mass in clusters is determined in two parts.  The first is a sum of cluster masses above a given ($\approx$completeness) limit.   We determine this completeness limit M$_{\rm lim}$ in each age interval to be the value where the mass function begins to fall significantly below an extrapolated power law, as shown in Figure~8 in Appendix~A.
The second part extrapolates a power-law mass function with an index of $\beta=-2$ below the completeness limit down to an assumed minimum cluster mass of $10^2~M_{\odot}$. This assumed shape is consistent with the observed cluster mass functions in our BCDs, which are presented and discussed in Appendix~A.
The total mass in stars, i.e. the denominator, is calculated as the elapsed time $\times$ the assumed star formation rate.
We assume the SFRs compiled in column~4 of Table~\ref{tab:sample}, and an uncertainty of $25$\%.
\par 
We assess if incompleteness significantly impacts our results for $\Gamma$.
The dominant  source of incompleteness is our inability to identify faint (i.e. lower mass) clusters in regions with high background.  Incompleteness will result in artificially low estimates for $\Gamma$ if we assume a too-low value for M$_{\rm lim}$, but $\Gamma$ should be stable for values of M$_{\rm lim}$ that are unaffected by incompleteness.  Therefore, a  simple test is to estimate $\Gamma$ using different values of M$_{\rm lim}$. 
We find that the $\Gamma$ estimates are nearly identical if we adopt values for M$_{\rm lim}$ higher than those shown in Figure~8, and drop if we adopt lower values, as expected.   Therefore, incompleteness does not adversely impact our estimates of $\Gamma$.  We also find that the specific apertures used for cluster photometry and to estimate the background have a negligible impact on our $\Gamma$ estimates.  If we adopt the age-dating results when a higher maximum E(B-V) value is allowed in the fits, $\Gamma$ increases to $\sim 26$\% and $\sim30$\% for ESO185-IG13 and Haro~11, respectively.

\par 
Our estimated values of $\Gamma_{\rm 1-10~Myr}$ range from $\approx15-39$\%, and are compiled in Table~\ref{tab:gamma}.  We find that $\Gamma_{\rm 1-10~Myr}$ estimated from a second, independent code returns nearly identical values.
We also compile the $\Gamma_{\rm 1-10~Myr}$ results reported by \citet{Adamo10,Adamo11b} for all three galaxies.  These are all systematically higher than our results, with $\Gamma_{\rm 1-10~Myr}=50$\% for ESO338 and Haro11 and $\Gamma\approx26$\% for ESO185. We discuss the likely reasons for the discrepancies in Section~4.4.
\par 
We use the same methodology to estimate the fraction of stars found in older, surviving clusters, and compile the results in Table~3.  Older clusters hold onto fewer stars, with $\Gamma_{\rm 10-100~Myr} \approx 5-8$\% and $\Gamma_{\rm 100-400~Myr} \approx 1-2$\%.  These results however, are less certain than those for $\Gamma_{\rm 1-10~Myr}$ because we do not have a way to directly determine the average SFR in these specific age intervals.

\begin{table}[ht]
	\caption{The Fraction of Stars in Clusters}\label{tab:gamma}
	\centering
	\begin{tabular}{lccccccc}
	    \hline\hline
		Galaxy & $\Gamma_{\rm 1-10~Myr}$  & $\Gamma_{\rm 1-10~Myr}$ & $\Gamma_{\rm 10-100~Myr}$ & $\Gamma_{\rm 100-400~Myr}$ &    \\
		Name & This Work & Adamo & This Work & This Work & \\ 
		 & (\%) & (\%) & (\%) & (\%)   \\ 		
		\hline \hline
ESO~185-IG13  &  $15\pm4$ & $26\pm5$ & 7.4 & 1.5 &  \\ \\
ESO~338-IG04  & $39\pm10$  & $50\pm10$ & 6.5 & 0.8 &   \\ 
Haro11 & $18\pm5$ & $50^{+13}_{-15}$ & 5 & 0.5 &   \\ 
        \hline
	\end{tabular}
	\end{table}

\subsection{How does $\Gamma$ Vary with the Intensity of Star Formation?}

\begin{figure}[!ht]
	\centering
\includegraphics[width=\textwidth]{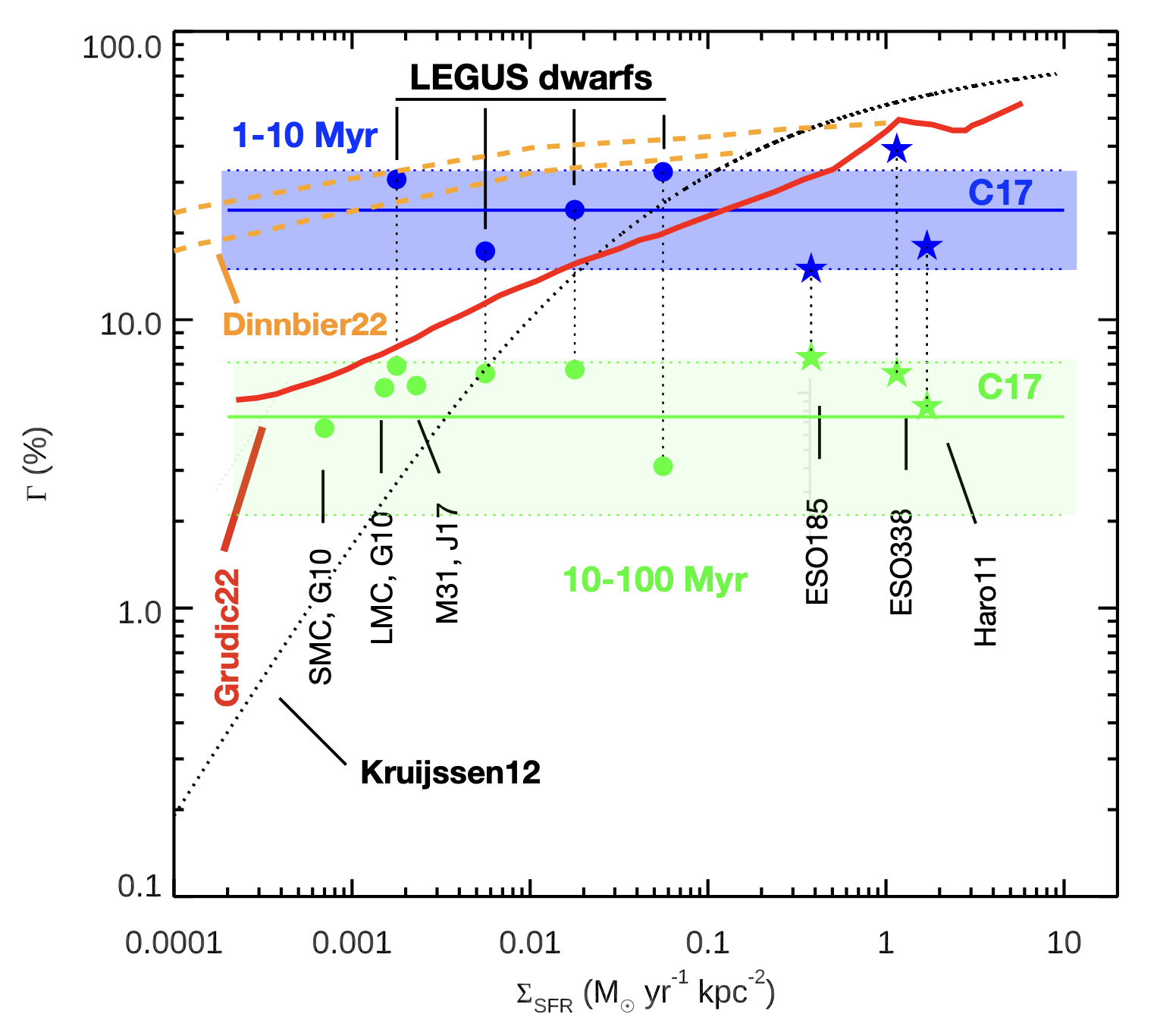}
	\caption{Our new estimates of $\Gamma_{\rm 1-10~Myr}$ and $\Gamma_{\rm 10-100~Myr}$ are shown by the blue and green stars, respectively. 
	The blue and green shaded regions show the mean and standard deviation for $\Gamma_{\rm 1-10~Myr}$ ($24\pm9$\%) and $\Gamma_{\rm 10-100~Myr}$ ($4.5\pm2.5$\%) from the eight galaxies studied by \citet{Chandar17}.  The filled circles show literature values for cluster formation efficiency, color-coded by age, determined for galaxies with a low $\Sigma_{\rm SFR}$.  Composite values of $\Gamma_{\rm 1-10~Myr}$ and $\Gamma_{\rm 10-100~Myr}$ estimated 
from 23 dwarf galaxies in 4 bins of $\Sigma_{\rm SFR}$ from the LEGUS survey 
(Cook et al. 2023)
 are included, along with $\Gamma_{\rm 10-100~Myr}$ estimates for M31 \cite{Johnson16}, and the Small \& Large Magellanic Cloud \citep{Goddard10}.} \label{fig:GammaLit}
\end{figure}

We show one of the main results of this work in Figure~\ref{fig:GammaLit}. Our new calculations are shown as the large blue ($\Gamma_{\rm 1-10~Myr}$) and green ($\Gamma_{\rm 10-100~Myr}$) star symbols, and plotted against $\Sigma_{\rm SFR}$.  The blue and green shaded regions show the mean and standard deviation of $\Gamma_{\rm 1-10~Myr}=24\pm9$\% and $\Gamma_{\rm 10-100~Myr}=4.5\pm2.5$\%, respectively, from the eight galaxies studied by \citet{Chandar17}. Our new results for the BCD galaxies fall within their respective age ranges within the errors.

Figure~\ref{fig:GammaLit} also plots results published for galaxies with lower $\Sigma_{\rm SFR}$.   
Cook et al. (submitted) recently grouped 23 nearby dwarf galaxies observed as part of the LEGUS project into four bins of $\Sigma_{\rm SFR}$, to improve cluster statistics.  We present their binned results based on SFR estimates derived from counts of young stars, stellar clusters, and associations \citep{Cignoni19}, and shown in Figure~\ref{fig:GammaLit} as blue ($\Gamma_{\rm 1-10~Myr}$) and green ($\Gamma_{\rm 10-100~Myr}$) circles. 
We also include commonly cited $\Gamma_{\rm 10-100~Myr}$ results for three other galaxies:
the LMC and SMC \citep{Goddard10},\footnote{\citet{Goddard10} found values of $5.8\pm0.5$\% (LMC) and $4.2^{+0.2}_{-0.3}$\% (SMC) for $\Gamma_{\rm 10-100~Myr}$, very similar to the calculations of $5^{+3}_{-1}$\% (LMC) and $3^{+2}_{-1}$\% (SMC) made by \citet{Chandar17} for the same age interval.}, and M31 \citep{Johnson16}.

Together, the galaxies in Figure~\ref{fig:GammaLit} cover a range of more than 3000 in $\Sigma_{\rm SFR}$, sufficiently large to draw conclusions about the relationship between $\Gamma_{\rm 1-10~Myr}$ (and $\Gamma_{\rm 10-100~Myr}$) versus $\Sigma_{\rm SFR}$. The best fit for each age interval is consistent with no correlation (slope of 0). This suggests there is {\em no significant trend in the cluster formation efficiency $\Gamma_{\rm 1-10~Myr}$ with $\Sigma_{\rm SFR}$ over the plotted range.}  $\Gamma_{\rm 10-100~Myr}$ also shows no correlation with $\Sigma_{\rm SFR}$.
\par 
Figure~\ref{fig:GammaLit} also includes three separate predictions that were shown in Figure~\ref{fig:GammaPredict}, which bracket the range and include the strongest correlation between $\Gamma-\Sigma_{\rm SFR}$ \citep{Kruijssen12}, a moderate one \citep{Grudic22}, and the weakest one \citep{Dinnbier22}.  The figure demonstrates that the weakest predicted correlation by \citep{Dinnbier22} best matches the observational results.
\par 
\subsection{Does the Fraction of Stars in Clusters Evolve Over Time?}
\begin{figure}[!ht]
	\centering
\includegraphics[width=\textwidth]{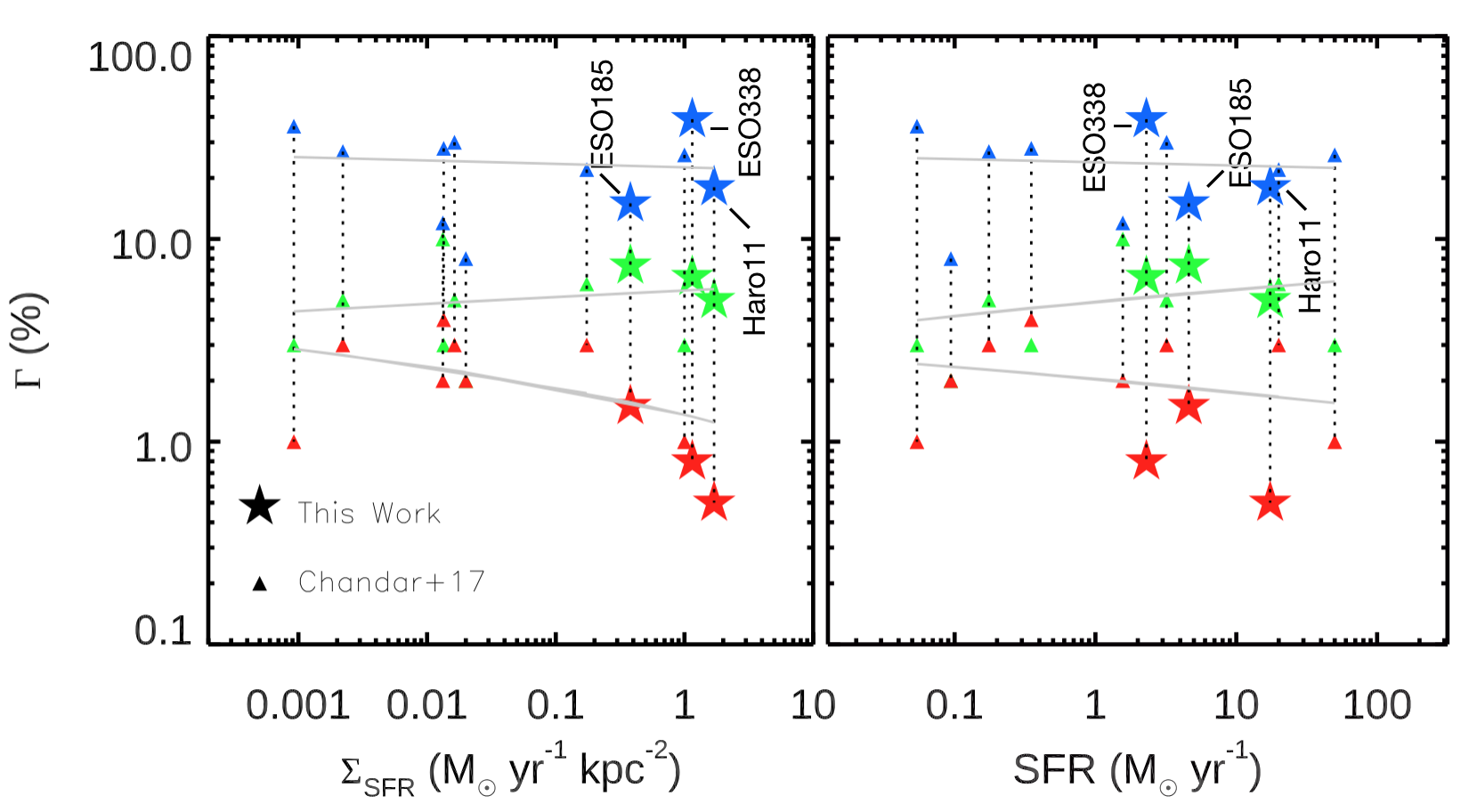}
	\caption{Estimates of $\Gamma_{\rm 1-10~Myr}$ (blue stars), $\Gamma_{\rm 10-100~Myr}$ (green stars), and $\Gamma_{\rm 100-400~Myr}$ (red stars) for the BCD galaxies ESO185-IG13, ESO338-IG04, and Haro11, studied here.  Results from Chandar et al. (2017) for 8 galaxies, including irregulars, spirals, dwarf starbursts, and on-going mergers, are shown as the triangles.  The main result of this figure is that 
	the fraction of stars in clusters declines as the clusters age at all $\Sigma_{\rm SFR}$. } \label{fig:Gamma}
\end{figure}

Figure~\ref{fig:Gamma} shows a second important result. 
Our calculations for ESO185, ESO338, and Haro11 suggest that the fraction of stars in clusters decreases from $\Gamma_{\rm 1-10~Myr}\approx 15-39$\% (blue stars) to $\Gamma_{\rm 10-100~Myr}\approx5-8$\% (green stars), and decreases again for $\Gamma_{\rm 100-400~Myr}\approx1-2$\% (red stars).
Results published previously by \citet{Chandar17} for 8 galaxies (LMC, SMC, NGC~4214, NGC~4449, M83, M51, the Antennae, and NGC~3256) show similar declines in the fraction of stars that remain in clusters over time.  These are plotted as the triangles in Figure~\ref{fig:Gamma}, and color-coded in the same way as for our BCD galaxies.

We plot the calculations for $\Gamma$ in all three age intervals versus $\Sigma_{\rm SFR}$ in the left panel, and versus SFR in the right panel.
 The best fit relation in each age range is shown by the gray lines in Figure~\ref{fig:Gamma}, with no obvious difference between $\Sigma_{\rm SFR}$ and SFR.  Each line is essentially flat, within the errors (i.e. has a slope consistent with 0). Our results indicate 
that the fraction of stars that remain in clusters decreases over time.
To date, few works (besides our own) have published results for $\Gamma_{\rm 100-400~Myr}$.  In fact, this is the first work to present calculations of $\Gamma_{\rm 100-400~Myr}$ for BCDs. It will be important for future works to include calculations in this age range to better disentangle cluster formation and disruption.

\subsection{Why We Find a Different $\Gamma - \Sigma_{\rm SFR}$ Trend than Previous Studies}

There are two primary reasons we find little-to-no correlation between $\Gamma$ and $\Sigma_{\rm SFR}$, while most previous observational works claim a strong correlation. 
The first is that previous works have plotted results from different age ranges for different galaxies, directly comparing $\Gamma_{\rm 10-100~Myr}$ for galaxies that have low $\Sigma_{\rm SFR}$, with $\Gamma_{\rm 1-10~Myr}$ for galaxies with high $\Sigma_{\rm SFR}$.
It is clear from Figure~\ref{fig:GammaLit} that a plot which included the green circles representing $\Gamma_{\rm 10-100~Myr}$ for the LMC, SMC, and M31, and also the blue stars representing $\Gamma_{\rm 1-10~Myr}$ for ESO185, ESO338, and Haro11, would roughly follow the trend of increasing $\Gamma$ with $\Sigma_{\rm SFR}$ predicted by \citet{Grudic22} and shown by the red line. This type of mixed-age comparison of observational results is biased and incorrect, yet has been presented in many works, including observation-based studies \citep[e.g.,][]{Goddard10,Adamo15,Adamo20,Johnson16,Messa18a} and theoretical ones \citep[e.g.,][]{Kruijssen12,Li17,Li18,Lahen19}. 
This bias was first pointed out by \citet{Chandar17}, and is easy to correct.
For $\Gamma$, going forward {\em only results calculated from the exact same age interval should be compared between galaxies.}

The second reason is more subtle yet fundamental, because it is related to correctly age-dating clusters.
To estimate the age, reddening, and mass of clusters in star-forming galaxies, it is standard to perform some type of SED-fitting, where measured broad-band colors are compared with predictions from stellar evolutionary models (similar to the method described in Section~3.1). The main challenge is to break the age-reddening degeneracy, since clusters can appear red because they are older or because they are reddened due to the presence of dust \citep[e.g.,][]{Adamo10,Calzetti15, Turner21}.
Usually there are 2 or 3 combinations of age and reddening which match the measured broad-band colors of a cluster (e.g., see Figure~4 in Whitmore et al., submitted).  Including information on whether or not line emission is present, from H$\alpha$ photometry for example, significantly improves age estimates by distinguishing between very young, H$\alpha$ emitting clusters and older ones which no longer show line emission \citep[e.g.,][]{Chandar10b,Fouesneau12,Ashworth17,Whitmore20}. 

A high maximum allowed E(B-V) in age-dating can lead to catastrophic age and mass determinations for some clusters when no narrow-band photometry is included (as described in detail by \citet{Whitmore20,Whitmore23}.
During the fitting procedure, the same range of reddening (from 0.0 up to some maximum E(B-V)) is typically adopted for all clusters, regardless of how much dust is nearby.  As discussed in Section~3.2, we believe that a number of older clusters in ESO185, ESO338, and Haro11 were fit with ages younger than log$\tau \lea 7.0$ by \citet{Adamo10,Adamo11a}  because they adopted a high maximum value of E(B-V)$=1.5$~mag ($A_V\sim 4.8$~mag for a Milky Way-like extinction curve) for all clusters but did not have H$\alpha$ photometry to help break the age-reddening degeneracy. 

We perform several experiments to assess the impact that catastrophic age-dating could have on $\Gamma_{\rm 1-10~Myr}$.  
For ESO185, Adamo et al. (2010) found ten very young $<10$~Myr clusters with masses $\geq 10^5~M_{\odot}$, while we found a single cluster that satisfies this age-mass criteria.  We find that our result for $\Gamma_{\rm 1-10~Myr}$ nearly doubles and is similar to that found by Adamo et al. (2011) if we artificially match the number (10) and approximate masses of their clusters in this age-mass regime.
For Haro11, if we  $\sim$double the mass of the most massive $<10$~Myr cluster to match that in Adamo et al. (2011), $\Gamma_{\rm 1-10~Myr}$ increases by $\sim10$\%.  If we assume the most (three most, five most) massive clusters with ages $\sim15-20$~Myr were age-dated to $\sim3-6$~Myr, we find $\Gamma_{\rm 1-10~Myr}$ increases by $\sim6$\% ($\sim15$\%, $\sim24$\%).  These differences can account for the offset between our results and Adamo et al.'s for $\Gamma_{\rm 1-10~Myr}$.  Our experiments establish that the systematic age-dating error that affected previous studies of the cluster populations in ESO185 and Haro11 (and likely ESO338) quite likely led to an artificial increase in $\Gamma_{\rm 1-10~Myr}$.  

\section{Summary and Conclusions}

Nearly all current simulations of cluster formation and early evolution predict that the fraction of stars that form in bound clusters increases with the amount of star formation per area or $\Sigma_{\rm SFR}$ in the host galaxy, although the exact relationship varies significantly between different works. 
However, these simulations rely on a large number of uncertain inputs and prescriptions, particularly for star formation and stellar feedback.
Because blue compact dwarf (BCD) galaxies have amongst the highest measured $\Sigma_{\rm SFR}$ within 100~Mpc, their cluster populations provide important input and constraints on simulations.

The fraction of stars formed in $1-10$~Myr clusters, or $\Gamma_{\rm 1-10~Myr}$, was previously estimated to be $\sim50$\% for the BCD galaxies ESO338 and Haro11, and $\sim26$\% for ESO185 \citep{Adamo10, Adamo11a}.  In this work, we found significantly lower values of $15\pm4$\% (ESO185), $39\pm10$\% (ESO338), and $18\pm5$\% (Haro11), based on deeper HST observations which include narrow-band H$\alpha$ imaging taken  as part of the Clusters, Clumps, Dust and Gas (CCDG) survey.  
We compared our new results with previous ones for galaxies with lower $\Sigma_{\rm SFR}$, and found that $\Gamma_{\rm 1-10~Myr}$ does {\em not} vary with $\Sigma_{\rm SFR}$, over a range of $\approx3000$ in $\Sigma_{\rm SFR}$, whereas current simulations predict an increase of factors between $\sim1.5$ and 100 over our observed range of $\Sigma_{\rm SFR}$.
We also found the fraction of stars that remain in older clusters drops to $\Gamma_{\rm 10-100~Myr}\approx5-8$\%,
and to $\Gamma_{\rm 100-400~Myr}\approx1-2$\% in our BCDs, as expected due to early cluster disruption.

We believe there are two reasons that our results
contradict previous works which have claimed a strong dependence of cluster formation efficiency on $\Sigma_{\rm SFR}$.
The first is that the previously reported increase was based on mixing results for $\Gamma_{\rm 1-10~Myr}$ for galaxies with higher $\Sigma_{\rm SFR}$, with $\Gamma_{\rm 10-100~Myr}$ for galaxies with lower $\Sigma_{\rm SFR}$.  Specifically, $\Gamma_{\rm 10-100~Myr}$ values determined for the SMC and LMC by \citet{Goddard10} and for M31 by \citet{Johnson17} have regularly been compared with $\Gamma_{\rm 1-10~Myr}$ estimates for galaxies like ESO185, ESO338, and Haro11.  
{\em It is critical that going forward, comparisons of $\Gamma$ between galaxies are based on the exact same age intervals.}

The second reason for the discrepancy is a key new finding of this work.  Previous age estimates for a number of older clusters in the BCD galaxies studied here appear to have been erroneously underestimated and best fit with 
very young $<10$~Myr ages and moderate reddening.  This age-dating error artificially boosted the mass and $\Gamma_{\rm 1-10~Myr}$ estimates.
We believe the age-dating concerns identified in this work may have affected a number of previous studies as well.

\par 
If confirmed, the observational results from this work present a challenge to most current theoretical models and simulations of cluster formation, because they
suggest that at least on galaxy-scales, outcomes of the star and cluster formation process do not vary strongly with the intensity of star formation.

\acknowledgements
R.C. acknowledges support from HST-GO-15469. We thank the anonymous referee for suggestions that improved our paper.
\bibliography{master}
\appendix 
\section{Cluster Mass Functions in Blue Compact Dwarfs}

A number of cosmological simulations that form entire cluster populations have measured the initial cluster mass function that is produced. 
Some of these simulations find an initial cluster mass function that is better represented by a Schechter function than by a power-law, with an upper mass cutoff $M_*$ that increases with $\Sigma_{\rm SFR}$ (e.g., Li \& Gnedin 2017, Li et al. 2018; Pfeffer et al. 2019). A Schechter-like shape suggests there is a physical upper mass with which clusters can form out of the ISM. 
Other simulations, for example from FIRE-2 (Grudic et al. 2022) create more complicated mass functions, which vary over time and by region and do not follow a simple power law and/or Schechter distribution. 

The observed shape of the cluster mass function in different galaxy environments, particularly whether or not there is a physical cutoff or truncation at the upper end, provides important constraints on the relationship between clusters and molecular clouds.
In this Appendix, we present the first assessment of the shapes of the cluster mass functions in ESO185-IG13, ESO338-IG04, and Haro11, and find that they are well-represented by a single power law, $dN/dM \propto M^{\beta}$, with $\beta \approx-2$. 

\subsection{Maximum Likelihood Fits}

\begin{figure}[!ht]
	\centering
\includegraphics[width=\textwidth]{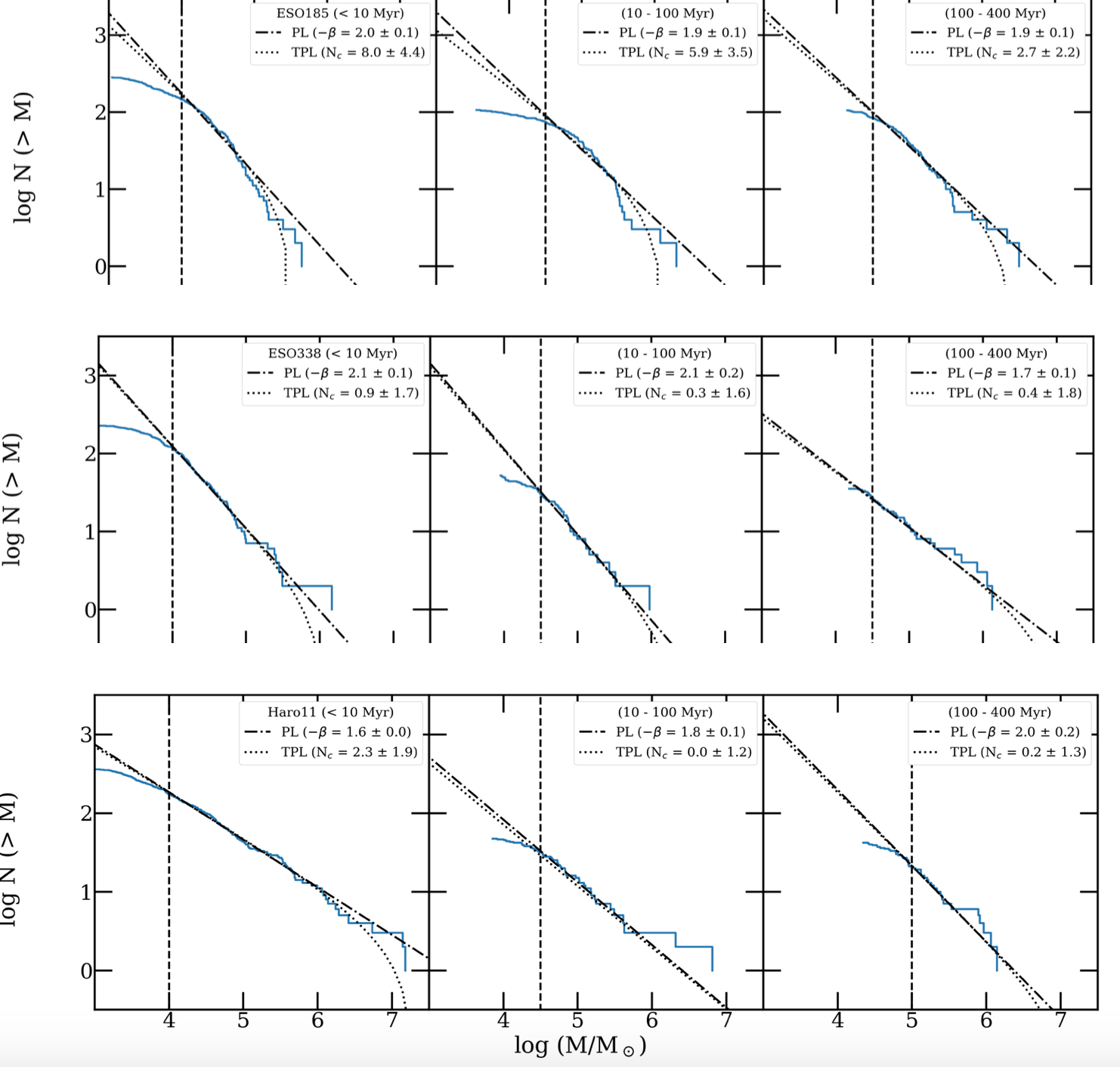}
	\caption{Cumulative mass functions are shown in blue for the indicated age intervals.  The best fit power-law is shown as the dashed-dotted line, and the best fit Schechter function in the dotted lines.  The best fit power law index $\beta$ and the statistic related to whether or not there is a truncation in the power-law are listed in each panel. The vertical dashed lines show the lower mass value used in the Maximum Likelihood analysis.} \label{fig:MF}
\end{figure}

In Figure~\ref{fig:MF}, we present the cumulative cluster mass functions in three age intervals: $1-10$~Myr (left), $10-100$~Myr (middle), and $100-400$~Myr (right), with results for ESO185 shown in the top panels, ESO338 in the middle, and Haro11 in the bottom panels.  We determine the completeness limit as the mass at which each distribution begins to fall significantly below an extrapolated power law (by a factor of $\sim$two; see Mok et al. 2019, 2021).
These lower mass limits are shown as dashed vertical lines, and adopted for the Maximum likelihood fits described below as well as the $\Gamma$ calculations presented in Section~4.1.  The best fit power law and truncated power law from the MSPECFIT code \citep{Rosolowsky05} are shown in each panel, as indicated.  The best fit power-law indices, $dN/dM \propto M^{\beta}$ are all between $-1.6$ and $-2.1$, with a median of $\beta=-1.9$.

We use the maximum-likelihood method described in \citet{Mok19} to determine the best-fit value and confidence intervals for $\beta$ and $M_*$ when fitting a Schechter function to the cluster masses above the completeness limit, $\psi(M) \propto M^{\beta}~\mbox{exp}(-M/M_*)$. This method does not use binned data, which can hide weak features at the ends of the distribution,  or cumulative distributions, where the data points are not independent of one another. We compute the likelihood $L(\beta,M_*)=\Pi_i P_i$ as a function of $\beta$ and $M_*$, where the probability $P_i$ for each cluster is given by:
\begin{equation}
P_i = \frac{\psi(M_i)}{\int_{M_{\rm min}}^{\infty} \psi(M)dM} 
\end{equation}
and the product is over all clusters above the completeness limit (see e.g. Chapter~15.2 of \citet{Mo10}).  We set the upper integration limit in equation (1) to be 100 times the mass of the most massive cluster, which is sufficient for convergence.  Next, we find the maximum-likelihood $L_{\rm max}$ using the \citet{Nelder65} method, and use the standard formula:
\begin{equation}
\ln L(\beta, M_*) = \ln L_{\rm max} - \frac{1}{2} \chi_p^2(k)
\end{equation}
where $\chi_p^2(k)$ is the chi-squared distribution with $k$ degrees of freedom at $p$ confidence level to determine the 1-, 2-, and 3-$\sigma$ confidence contours.
\par
\begin{figure}[!ht]
	\centering
\includegraphics[width=\textwidth]{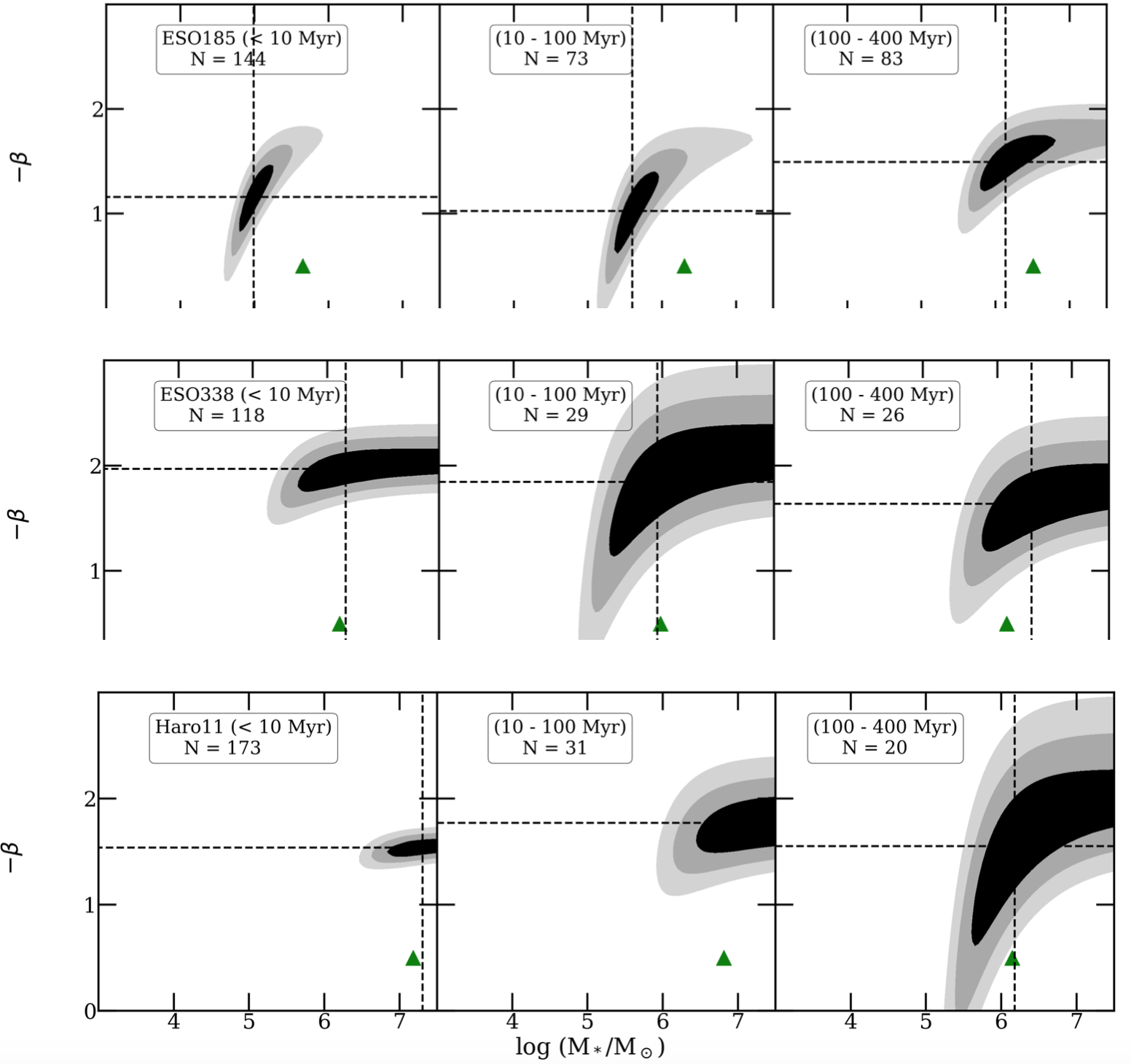}
	\caption{Likelihood fits of the Schechter parameters $\beta$ and $M_*$ to cluster masses in three age intervals, $<10$ Myr (left), 10–100 Myr (center), and 100–400 Myr (right). Results for ESO185 are in the top panels, ESO338 in the middle, and Haro11 in the bottom panels.  The dashed lines show the best-fit values of $\beta$ and $M_*$, while the boundaries of the shaded regions show the $1\sigma$, $2\sigma$, and $3\sigma$ confidence contours.  The green triangles show the maximum cluster mass in each sample.} \label{fig:ML}
\end{figure}

 
Figure~\ref{fig:ML} shows the best-fit values of $\beta$ and $M_*$ (dashed lines) that result from our Maximum Likelihood fitting analysis.  The shaded regions show the 1, 2, and 3$\sigma$ contours. The contours show a short diagonal portion,  which indicates the trade-off between a steeper value of $\beta$ and a higher cutoff mass $M_*$, and then a relatively flat portion up to the highest mass tested.  The green triangle shows the most massive cluster in each sample.  We do not include uncertainties in the cluster mass estimates in the fits, which would serve to further increase the uncertainties in the Schechter parameters $\beta$ and $M_*$.  

\subsection{Implications for the Upper End of the Cluster Mass Function}

We are interested in answering the question:
is there a physical mass limit with which clusters can form in ESO185, ESO338, and Haro11, and if so, can we detect it?  As in our previous studies, we performed  fits to the mass function of clusters in the 1-10~Myr, 10-100~Myr, and 100-400~Myr age intervals.  Ideally, we should find consistent estimates of, or limits on, the Schechter parameters $\beta$ and $M_*$ in all three intervals, since the physics of cluster formation are unlikely to change significantly over the relatively short period of $4\times10^8$~yr covered by our study. It is physically implausible for the upper cutoff $M_*$ to increase with age, but quite possible that systematic errors could affect the fit results.
The 100-400~Myr interval covers the longest elapsed time, so for a given SFR, this interval is likely to be the most reliable for determining $\beta$ and $M_*$ (Mok et al. 2019).  Haro11 is an exception, since the most massive clusters (and likely most intense star formation) occurred $\approx15-30$~Myr ago in this galaxy, with very few clusters with ages between 100 and 400~Myr detected.

None of the fits show evidence for a specific upper mass cutoff in ESO338 and Haro11, since the horizontal confidence contours continue without closing to the right edge of the diagrams.
This means that the cluster masses are {\em consistent with being drawn from a single power law}, but an upper cutoff (over this range of $M_*$) cannot be ruled out.
These results indicate that the value of $M_*$ returned by the maximum likelihood fit is indeterminant and represents a lower limit to the actual value. For ESO185, the two younger age intervals show weak evidence for a cutoff $M_*$ with a very shallow estimate of $\beta \approx-1$; this shallow value of $\beta$ reflects correlations between the exponent $\beta$ and cutoff $M_*$, since there is a trade-off between a steeper value of $\beta$ and higher value of $M_*$.  Taken at face value, the best fit value of $M_*$ increases with age, which is unlikely to be a physical result.  
{\em We conclude that the three BCD galaxies studied here have cluster masses that are consistent with being drawn from a pure power law, with a wide range of permitted cutoff masses from $M_*\approx10^6~M_{\odot}$ to $\infty$.}

\end{document}